# Accelerated Insulation Aging Due to Fast, Repetitive Voltages: A Review Identifying Challenges and Future Research Needs

Mona Ghassemi
Virginia Polytechnic Institute and State University
The Bradley Department of Electrical and Computer Engineering
Blacksburg, VA 24061, USA

## ABSTRACT

Although the adverse effects of using power electronic conversion on the insulation systems used in different apparatuses have been investigated, they are limited to low slew rates and repetitions. These results cannot be used for next-generation wide bandgap (WBG)-based conversion systems targeted to be fast (with a $dv/dt$ up to 100 kV/µs) and operate at a high switching frequency up to 500 kHz. Frequency and slew rate are two of the most important factors of a voltage pulse, influencing the level of degradation of the insulation systems that are exposed to such voltage pulses. The paper reviews challenges concerning insulation degradation when benefitting from WBG-based conversion systems with the mentioned $dv/dt$ and switching frequency values and identifies technical gaps and future research needs. The paper provides a framework for future research in dielectrics and electrical insulation design for systems under fast, repetitive voltage pluses originated by WBG-based conversion systems.

Index Terms — accelerated aging, insulation systems, fast, repetitive voltage pulses, wide bandgap-based conversion systems

## 1 INTRODUCTION

**ACCELERATED** aging of insulation systems envisaged for the next-generation WBG-based power electronics building blocks (PEBB) and accelerated aging of existing insulation systems used in all other power system apparatuses such as rotating machines, cables and cable terminations, transformers, etc. exposed to fast, repetitive voltage pulses originated by WBG-conversion systems is the most significant barrier to realizing high-voltage, high-power density conversion devices and systems as well as benefitting from them in power systems for various applications.

Although the adverse effects of using power electronic conversion on the insulation systems used in the apparatuses mentioned above have been investigated, they are limited to low slew rates and repetitions. Frequency and slew rate are two of the most important factors of a voltage pulse, influencing the level of degradation of the insulation systems that are exposed to such voltage pulses. There are no experimental data for insulation degradation under voltage pulses with high slew rates of up to 100 kV/µs and high repetition rates (frequency) up to 500 kHz targeted by US military and in the near feature for power grid applications.

A modular design concept (namely PEBB) [1], initiated in late 1994 by the U.S. Office of Naval Research (ONR) [2, 3], was envisioned for converters. PEBB incorporates progressive integration of power devices, gate drives, programmable processors, and other components into building blocks with specific functionalities and interfaces to serve multiple applications [4, 5], resulting in reductions in cost, loss, size, and engineering effort for the application and maintenance of power electronics systems [4]. The performance of PEBBs can be improved by using silicon carbide (SiC), as the most promising WGB material for high voltage high power density applications, devices instead of Silicon (Si) devices. The main advantage of SiC is that it features a breakdown electric field nearly an order of magnitude higher than Si, as well as being capable of exceedingly fast commutation while exhibiting reduced conduction and switching losses and being able to operate at higher temperatures [6-8].

Rated at 100 kW, 1 kV DC bus, and 100 kHz switching frequency, the PEBB 1000, developed by Virginia Tech's Center for Power Electronics Systems (CPES) for ONR, contains two 1.7 kV, 300-A SiC MOSFET phase-leg modules in an H-bridge configuration [9, 10]. Using the knowledge and experience gathered during development of the PEBB 1000, CPES began the PEBB 6000 program, a full-bridge, 10 kV, 240 A SiC MOSFET-based building block rated at 1 MW, 10 MW/m³, 6 kV DC bus, and operating at 20 kHz, in early 2018. The voltage waveforms of both the PEBB 1000 and the PEBB 6000 have a high slew rate of 50 kV/µs. In this regard, even though the highest switching frequency recorded in US Department of Defense (DoD)'s sponsored high voltage, high power SiC-based PEBB development program is 100 kHz, higher equivalent switching frequencies via connecting PEBBs in series or parallel can be envisaged.

As it is reviewed and discussed in this paper, insulation systems of different power system apparatuses can adversely be affected by fast (50 kV/µs), repetitive (100 kHz) voltage pulses generated by the new, state-of-the-art PEBBs mentioned above and it gets worse for the targeted values. The technical gap is that the slew rate and frequency of experimental investigations reported so far are limited to the values that are less than those achieved (PEBB 1000) and targeted for next-generation WBG-based converters. Moreover, neither in-depth (physical-based) explanation nor a model is presented for the test results. In other words, although WBG devices are revolutionizing power electronics, electrical insulating systems have not been prepared for such a revolution. In this paper, technical gaps in different parts of insulation systems affected by WBG-based devices are identified and discussed. This study will also provide a useful framework and point of reference for the future development of electrical insulation materials and systems to address the accelerated aging issue under fast, repetitive voltage pulses not experienced so far.

The arrangement of the sections of this paper is based on the envisaged sequence of future research needs as follows. First insulation degradation under fast, repetitive voltage pulses should experimentally be investigated due to partial discharge (PD) and water treeing. This is discussed in Section 2. In this regard, either special testing electrode geometries may be designed or conventional testing component types can be used. For example, for the later for motor stator winding, (I) twisted pairs simulating turn/turn insulation systems in wire wound machines, (II) motorettes simulating all parts of the insulation system, i.e., the turn/turn, phase/ground, and phase/phase of a random wound machine, (III) formettes simulating groundwall insulation and the stress grading system, and (IV) turn/turn samples that are the equivalent of a twisted pair for a Type II system can be used for tests.

Note that test results are electrode geometry and test condition-dependent. The development of physically based "theory" model(s) validated by experimental results can address the mentioned issue. Having such models, simulations can be carried out for realistic geometries and multi-stress conditions. This is discussed in Section 3.

However, the question that arises is "What is a realistic voltage waveform that insulation systems in a power system is exposed to?" Due to two phenomena, 1) impedance mismatch and 2) uneven voltage distribution caused by high $dv/dt$, the voltage distribution on different insulation systems, such as motor stator winding, stress grading system in form-wound machines, cables and, cable termination is entirely different from the voltage pulses generated in converter terminals. To obtain these voltage distributions, electromagnetic transient (EMT) models of different apparatuses should be created. It is discussed in Section 4. For voltage pulses with a high slew rate of 100 kV/µs and a repetition rate of 500 kHz, very high frequencies (up to a few or a few ten MHz) for EMT models of the components are needed.

In this paper, the published investigations dealing with insulation degradation under fast, repetitive voltage pulses have been reviewed. In this regard, all reviewed papers reporting experimental investigations used applied voltage levels of higher than 1 kV for test setups.

## 2 INSULATION DEGRADATION DUE TO PD AND WATER TREEING IN SOLID DIELECTRICS, AND INSULATION DEGRADATION IN GELS: EXPERIMENTAL INVESTIGATIONS

Insulation degradation factors under fast, repetitive voltage pulses can be categorized into (1) space charge, (2) intrinsic aging, (3) PDs, and (4) treeing inception [11]. Space charge can be injected in an insulation system only at low frequencies, and it can be neglected for frequencies above 50 Hz. Experimental investigations in [12] show that the lifetime of twisted pairs immersed in silicone oil to prevent PD inception, significantly decreases under a high frequency (10 kHz), sinusoidal waveform compared with a 50 Hz sinusoidal waveform (for example 2 hrs; 400 hrs under 5.5 kVrms), attributing to intrinsic aging. However, compared to PD activity, intrinsic aging has only a minor effect. Here, the technical gap is that the frequency of the test voltage is limited to 10 kHz and a sinusoidal (not square) waveform.

### 2.1 INSULATION DEGRADATION DUE TO PD IN SOLID INSULATION

Under repetitive surges, PDs occur on each rising and falling flank of the voltage waveform. It should be noted that the PD is incepted only when a starting electron becomes available. Thus, there is a statistical delay between the moment when the voltage is sufficient for PD inception and the moment when the PD is incepted. The shorter the rise time, the larger the PD magnitudes [13]; thus, the shorter lifetime [14-18] as shown in Figure 1. Furthermore, it can be seen that lifetime dramatically decreases with increasing frequency. To address this issue, filters were proposed to increase the rise time of the surges impinging on the insulation system terminations [19, 20]. It should be noted that increasing the switching frequency has the advantage that the size and weight of the passive components such as output sine wave filters can be decreased [21, 22]. Sine wave filters are usually used to eliminate switching frequency harmonic contents at the output of the variable frequency drive (VFD), minimizing excess heat and voltage stress and insulation failures in the motor.

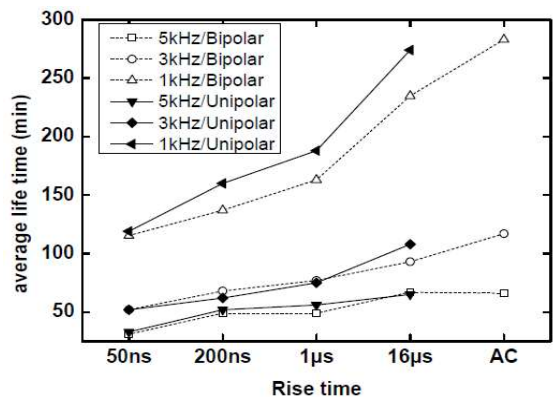

**Figure 1.** Average lifetime of single-contact point crossed enameled wire pairs as a function of frequency and rise time [14].

Here, the technical gap is that the slew rate and frequency of test voltages are limited to 70 kV/μs and 5 kHz. Although a high slew rate test voltage was used, the peak-to-peak of test voltages ($V_{pp}$) is under one value (for example 2.5 kV in [15] and 3.5 kV in [16]). In other words, we do not know what happens if tests are carried out for a high slew rate at different values of $V_{pp}$ (the influence of high field conditions on insulation aging has not been studied). Furthermore, neither in-depth (physical-based) explanation nor a model is presented for the test results [13-18].

## 2.2 INSULATION DEGRADATION DUE TO WATER TREEING IN SOLID INSULATION

Water treeing is one of the most significant reasons for degradation in polymeric cable insulation, and since water tree growth depends on the number of zero-crossings of the applied voltage [23-29], surge trains with high repetition rates lead to accelerated water tree growth. However, tests in [23-29] were carried out under sinusoidal waveforms (not square waveforms) that include superposition of a high-frequency 2 or 4-kHz AC voltage to various voltages such as dc, low-frequency (0.1 to 5 Hz) and power-frequency voltages. It was shown in [30, 31] that the length $L$ of water trees varies linearly when plotted as $\log L$ vs. $\log N$, where $N$ is the number of field cycles. This relation is nearly independent of the applied field, suggesting a fatigue-like process. Ion concentration has only a slight influence on water tree length, especially after long aging (>$1 \times 10^8$ cycles). Also, it was reported that the $\log L$ vs. $\log N$ relationship deviates from its linear behavior between 30 and 447 kHz [30, 31] as shown in Figure 2, supporting the idea that there is an upper-frequency limit where water treeing would be independent of frequency.

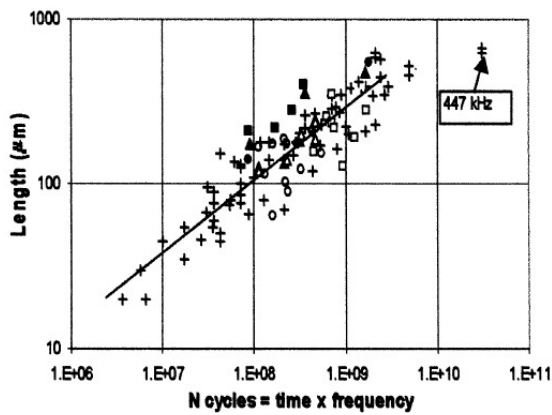

**Figure 2.** Water-tree length in PE soaked in NaCl at 22°C as a function of the number of field cycles with three different testing cells. Field ranging from few kV/mm up to ~100 kV/mm. Frequencies from 50 Hz to 447 kHz [33].

A model for water treeing for 50 Hz to 30 kHz AC fields, assumes that the field-induced stress, applied on nanocavities filled with a liquid, is larger than the yield strength of the polymer was introduced in [31-33]; thus, bonds will be broken, and the nanocavity will expand.

According to the model an equation was introduced as $L \approx \left(N\epsilon'\epsilon_0 n_0 t^{1/2} v_0 F^2/Y\right)^{1/3}$ where $L$ is water-tree length, $N$ is the number of cycles, $\epsilon'$ and $\epsilon_0$ are the dielectric constant of the liquid ($\epsilon' = 80$ for water) and the free space permittivity, respectively, $n_0$ is the initial diffusion concentration ($n_0 = 1.44 \times 10^4$), $t$ is the time, $v_0$ is the average volume of the free voids, $F$ is the electric field, and $Y$ is the yield strength (~$1.5 \times 10^7$ $N/m^2$ for PE at 22°C). Water treeing in XLPE insulation at a combined DC and high-frequency AC stress (DC stress +10% sinusoidal voltage at 5 kHz) was also investigated in [34]. Technical gaps regarding the research carried out so far are (1) the water tree models developed are valid up to 30 kHz and (2) experimental tests were carried out under sinusoidal waveforms (not square waveforms).

## 2.3 INSULATION DEGRADATION IN GELS

The first apparatus affected by insulation degradation under fast, repetitive voltage pulses is the apparatus producing such pulses, the PEBBs, and, in particular, WBG power modules. The silicone gel used in WBG power modules for encapsulation is prone to degradation. However, almost all research reported in the literature in this regard is about Si-IGBT. A thorough and in-depth review of PD measurements, failure analysis, and control in high-power IGBT modules has been done in [35, 36].

PD detection in power electronic modules based on phase-resolved partial discharge (PRPD) patterns was reported in [37-50]. The hypotheses proposed about the origin(s) of PDs in the module based on the measured PRPD patterns are not consistent and, in some cases, are even contradictory [41, 48]. The optical technique is a promising method to localize PDs accurately in a power electronics module [38, 45, 46, 48, 50, 51]. In [50, 51], results concerning both electrical and optical detection of PDs occurring in the silicone gel were presented. That work showed that optical measurements could be used to study PDs in transparent gels with any voltage shape and with very high sensitivity (<1 pC). However, PD in a power electronic module is an ultra-low-light-level phenomenon and a very compact intensified CCD camera is needed to record the small light intensities emitted by electroluminescence effects as well as light caused by PD [38]. Gel features are reminiscent of both liquids and solids: they have an elastic behavior due to their chemical cross-links as well as some self-healing properties. Compared to liquids, gels do not require watertight packaging, which greatly contributes to simplifying the design and reducing the cost. If the mechanism of electrical failure in solid insulation is still not fully understood comparatively, we know almost nothing for gels, such as silicone gels, used for power electronics modules.

When increasing blocking voltage up to 15-20 kV or more and switching frequency up to 500 kHz or more (into the MHz range) for the envisaged WBG (both GaN and SiC MOSFET) modules, a compact design creates an extremely poor environment for the silicone gel. This can get worse if a high-power capability is also added to the module where, in addition to the electrical stress, thermal stress is also added. The Achilles heel of this technology is accelerated insulation aging of the encapsulant material. To the best of our knowledge, no basic research has been conducted to address this issue. We know nothing about the mechanisms and phenomena leading to electrical failure in silicone gels subjected to fast, repetitive voltage pulses.

Moreover, geometrical techniques [52] or applying nonlinear field dependent conductivity materials on high field regions [53-55] for electric field control in power electronic modules are only based on simulation results and have not been implemented in practical modules.

## 3 MULTIPHYSICS MODELING

Teaching and research in electrical insulation and high-voltage technology have relied mainly on experimental techniques. However, instead of realistic geometries which may be expensive, uniform field gaps, e.g., Rogowski electrode or nonuniform field gaps, e.g., rod-rod gaps and rod-plane gaps, have often been used for experiments. These methods may lead to unreliable conclusions and, thus, an overdesign of electrical insulation systems is considered. Moreover, experimental results are described qualitatively instead of quantitatively. Some mechanisms are emphasized while other mechanisms are ignored or downplayed to enable justification of experimental results. Computer (numerical) modeling has also been limited to electric field calculation (single physics) without dealing with the underlying phenomena behind pre-breakdown in insulation materials.

A physically based "theory" approach is needed for the optimal design of electrical insulation systems. In a physically based "theory" approach, all physical mechanisms behind pre-breakdown phenomena are considered. The future of electrical insulation design belongs to a physically based "theory" approach which leads to reliable, as well as compact, designs. Rare efforts have been reported in the literature that use a physically based "theory" approach. Difficulties in a physically based "theory" approach are complex theoretical concepts, complex mathematical equations describing theories, numerical issues to solve non-linear and extremely coupled equations, and the need for high-performance computers with substantial memory resources to run and save simulations in a reasonable amount of computation time. Presently, researchers examine different materials and fillers in extremely time-consuming and expensive tests to explore a proper dielectric or optimal percent of a filler for a particular application. Many of these efforts fail since they are examining only one stress condition while there is a multi-stress (electrical, thermal, mechanical and environmental) condition in realistic situations. Through Multiphysics modeling based on a physically based "theory" approach, optimal designs can be achieved. The author has established this approach in different applications [56-65] including gas-solid and liquid-solid combination insulations.

However, such work still has not been done for a solid insulator with gas voids. We do not know what the equations and their associated parameters for solid insulation are. That will be a significant breakthrough in the modeling of solid insulation breakdown. Besides solid insulation, such modeling for gels, which have the properties of both solid and liquid insulation, is more difficult than solid insulation.

To the best of my knowledge, aging mechanism of solid insulations materials under fast, repetitive voltage pulses has been investigated in only one paper [66] where it is related to thermal processes triggered by losses generated in insulating materials due to damped oscillations of molecular dipoles in high-frequency electrical fields.

## 4 HIGH-FREQUENCY EMT MODELS

Voltage surges with steep fronts caused by turning semiconductor switches on/off in converters, travel through cables (in a shipboard application) and are reflected at interfaces due to impedance mismatches, giving rise to local overvoltages as shown in Figure 3. Phenomena that are typically associated with these repetitive overvoltages are partial discharges (PD) and heating in insulation systems, which contribute to insulation system degradation. Overvoltages may also lead to a dielectric breakdown.

IEC 60071-1:2006 [67] defines three parameters (pulse magnitude, front time, and half value time) to describe the waveshape of lightning, switching, and very fast transients (VFT). VFTs created by power converters on motor terminals usually have a waveform, as shown in Figure 3, that cannot be described by IEC 60071-1:2006 due to (1) reflections and (2) uneven distribution of turn voltages. To address this, IEC 60034-18-41 [68] and IEC 60034-18-42 [69] give a more descriptive waveshape. Reflections at cable/motor and cable/converter interfaces caused by impedance mismatches create a voltage overshoot at the leading edge of the pulse, increasing the risk of electrical breakdown. Figure 4 shows the size of the overshoot as a function of cable length for different rise times for a 2-level converter [70]. The curve to the left is for an impulse rise time of 50 ns, and the overshoot reaches a factor of 2 at a cable length of 4 m. Steep-fronted switching surges can cause undesirable transient voltage distributions as well as hot spots in the machine stator winding, the SG system, and cable terminations. These overvoltage stresses can cause premature deterioration or even failure of the insulation, resulting in a forced outage of the drive system [71].

Due to gassing problems and premature insulation failures in transformers subjected to high $dv/dt$ pulse voltages caused by converter switching, or circuit breaker switching (especially for wind turbine converter-fed transformers [72-74]), the effect of square voltage waveforms on transformer insulation materials has been investigated [73-75]. For oil-impregnated paper placed between rod-plane electrodes, the PD gas inception voltage, which is based on the dissolved gas analysis (DGA), depends on both the frequency and the rise time of the test voltage [74]. A decreasing trend in PDIV after 1 kHz was also reported in [75]. No test result has been reported over 10 kHz yet.

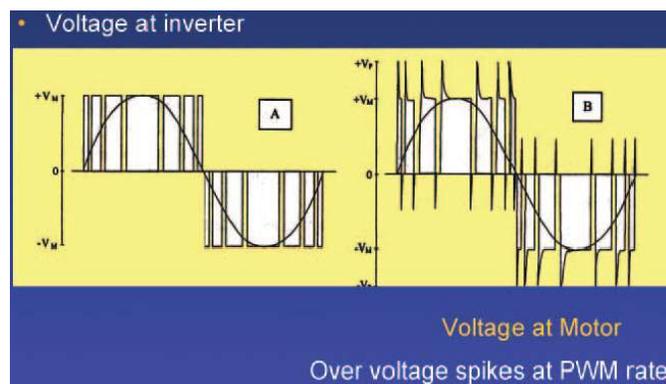

**Figure 3.** Comparison of voltages at a converter (A) and motor (B) terminals, b) Comparison of phase/phase, phase/ground and turn/turn voltages for a 2-level converter [70].

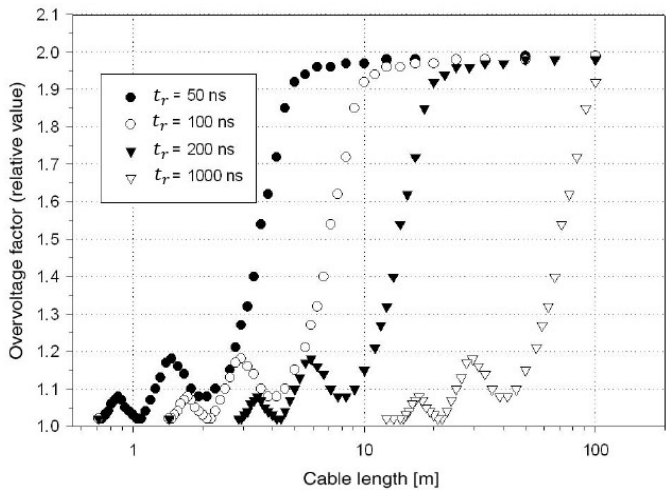

**Figure 4.** Voltage enhancement at the motor terminals as a function of cable length for various rise times (2-level converter) [70].

## 4.1 HIGH-FREQUENCY EMT MODELS FOR MACHINE STATOR WINDING

Various models for simulating the voltage distribution among the turns of different machine winding types have been presented. Initiated by [76-80], the multiconductor transmission line (MTL) model, which considers wave propagation within coils, has been developed as a lossy MTL model [81], non-linear MTL model [82], a finite element method (FEM) [81-83] or certain average techniques [84-86] to characterize different elements from the winding within MTL theory, and a frequency-domain methodology to calculate the distributed parameters of equivalent circuits [87] to predict the transient voltage distribution in a machine winding under steep-fronted surges. In this regard, it is well-known that the maximum voltage on the coil increases as the surge-front time decreases. However, in case studies, the highest slew rate and switching frequency are limited to 68 kV/$\mu$s [81] and 20 kHz, respectively. The challenges are as follows:

- For a SiC-based PEBB converter with a switching frequency of 500 kHz and a slew rate of 100 kV/$\mu$s and depending on the length and parameters of the cable on which the wave is traveling, the steep-fronted surge voltages at the motor terminals can have significant frequency components up to a few ten MHz or even a few hundred MHz. This means that a wideband model for motor stator winding needs to be extended one to two orders of magnitude higher than the frequency spectrum of the surge (in GHz range). However, a higher order model usually leads to large passivity violations and, thus, complex methods such as infeasible-interior point primal-dual methods [88] should be employed for passivity enforcement.
- The elements of the capacitance and inductance matrices for the overhang and slot regions can be calculated by the FEM-based COMSOL Multiphysics tool. The challenge here is that at high frequencies (MHz), the slot walls behave like a magnetic insulating wall due to eddy currents. Methods such as a multilayer method of images [89, 90] or analytical solutions [91, 92] are needed to calculate the inductance matrix. However, all these models are for up to a few MHz and, to the best of our knowledge, no study has been conducted for very high frequencies in the few hundred MHz or GHz ranges. Thus, either novel models should be developed, or required considerations should be identified for FEM-based modeling in COMSOL to address this difficulty.
- The series losses in the coil due to skin and proximity effects as described in [93] for transformer winding as well as the dielectric losses in the coil should also be calculated and added to the inductance and capacitance matrices obtained from above to obtain the series impedance and shunt admittance matrices required by the model. Both skin and proximity effects are frequency-dependent phenomena and are more critical in high-frequency ranges. However, they have only been modeled for up to a few MHz [87], and remain a challenge for higher frequencies.
- To obtain an accurate solution, cables and converters should be added into the model and a wideband model for a converter-cable-motor set is needed to be developed. In a Si-IGBT voltage source converter-based back-to-back (BtB) tie installation with a switching frequency of 1.26 kHz at the Eagle Pass substation in the State of Texas, a high-frequency component of around ten times the switching frequency (12.4 kHz) with an amplitude varied between 13 and 40% of the power frequency (60 Hz) voltage, depending on the operating mode of the BtB and the angle difference between the U.S. and Mexican grid were measured in compact type cable terminations rated at 24 kV with resistive/refractive stress grading (SG), led to hot spots in SG and eventually failure of the BtB [94]. This is the example that can be imagined for a 500 kHz switching frequency that high-frequency components in the MHz or GHz range may exist in the motor terminal as discussed above. Although polyethylene (PE), ethylene propylene rubber (EPR), and cross-linked polyethylene (XLPE) do not show remarkable changes in their complex permittivity within the range of frequencies caused by traditional converters, such as the Eagle Pass BtB tie converter mentioned above, this may not be the case for the SiC-based converters operating at 500 kHz, which is envisaged for future applications. In these cases (as a technical gap), frequency-dependent complex permittivity [95-98] should be taken into account for wideband modeling of cables. Some papers investigated cable-motor sets and, therefore, included impedance mismatch issues in uneven distribution voltage issues due to fast-fronted surges [84, 99, 100]. In all cases, the studied switching frequency of 20 kHz is too low to cause an issue. In other words, for a switching frequency of 20 kHz, there is sufficient time for damping of the surge propagating along the cable before initiation of the next switching operation in the converter. However, this is not the case for SiC-based PEBB converters, which are envisioned to operate at 500 kHz with a slew rate of up to 100 kV/$\mu$s at a converter that, in Navy ships, aircraft, micro and nano grids, is connected to motors with very short cable lengths. To the best of our knowledge, such critical situations have not been studied so far.
- The converter is assumed to be short-circuited or often not considered at all. This is because the cable or transmission line length connected to the other side of the converter is high and, in a study period, we do not have the wave reflecting from the far end. However, this may not be the case for ships, aircraft, micro and nano grids power system architectures where cable lengths are short. Therefore, the study zone should be expanded to include the DC or AC bus, other converters, and even, the whole power distribution system. A high switching frequency of 500 kHz and many impedance mismatch points will cause complex voltage waveforms and distributions on motor stator winding. Analysis of these waveforms and distributions becomes more difficult if the simultaneous operation of converters, under realistic conditions is considered.

After developing high frequency EMT models for all apparatuses of a power system benefitting from WBG-based conversion systems for various applications such as the medium voltage direct current (MVDC) system for future U.S. naval ships [101, 102], more-electric commercial aircraft, and all/hybrid electric aircraft concept introduced by NASA [103], all electric ground transportation, electrification of ultradeep water oil and gas fields [104], and on a larger scale, for DC power system distributions (DC microgrids [104]), EMT studies for the system will determine the voltage distribution and waveform on different insulation systems. Considering that voltage waveform as the input for the physically based "theory" model(s) discussed in Section 2 and validated with experimental results discussed in Section 1, this question can be answered that if we have a breakdown or unacceptable insulation degradation. If a breakdown or unacceptable insulation degradation occurs, various methods, including 1) mitigation methods, 2) changing the dimensions of insulation systems, 3) changing the insulating materials, 4) changing the specifications of converters, and 5) changing cable/transmission line lengths and specifications if possible, etc. can be considered. Then the EMT studies should be repeated or an optimization algorithm can be included, to achieve optimal electrical insulation design for that power system. It should be noted that for some components such as SG systems used in rotating machines and cable terminations, a coupled EMT-thermal model should be developed to determine temperature rise and hot spots there.

Although the challenges mentioned above still have not been addressed when using WBG-based conversion systems, the similar situations concerning transformer resonant overvoltages caused by cable-transformer interaction have been reported for conventional power systems where (short length) cable-transformer (motor) interaction can cause very high overvoltages. Switching transients, especially for wind applications subjected to more frequent switching operations and earth fault are two transient phenomena that can lead to resonant overvoltages at the LV terminal of a transformer as well as inside HV and LV windings [105-111]. The increasing number of transformer dielectric failures led to the CIGRE Working Group A2/C4.39 initiation for the computational assessment of impinging overvoltages on transformer terminals and internal stresses. Both laboratory tests and simulation results [105] showed that an 11 kV/230 V 300 kVA distribution transformer whose high-voltage side connected to a 27 m cable could experience a very high overvoltage (24 p.u.) on the open low-voltage side for a step voltage excitation on the cable.

Simulations showed that overvoltages as high as 43 p.u. could occur with the most unfavorable cable length (for a 20 m cable) [105]. The highest (resonant) overvoltages occur when the dominating frequency component of cable voltage matches one of the dominating frequency components of voltage transfer from the high-voltage side to the low-voltage side. As another example, the voltage transfer ratio for three winding designs: layer, disc, and pancake, of an 11/0.24 kV 500 kVA transformer was measured in [109]. All three windings had a dominant resonant frequency around 800 kHz. However, layer winding had the most dominant resonant frequency at 1.6 MHz and some resonant frequencies in a higher range.

## 4.2 THERMAL MODELING OF STRESS GRADING (SG) SYSTEMS

Besides insulation issues in the turn/turn, phase/ground, and phase/phase sections of rotating machines under fast, repetitive voltage pulses originating from converter-fed drives, the stress grading (SG) system in a form-wound machine is another part that is adversely affected from such surges [112-119]. As shown in Figures 5a and 5b, the SG system consists of a corona armor tape (CAT) and stress grading tape (SGT). This grading system works efficiently at power (50 or 60 Hz) frequency to relieve the very high field, which would occur at the edge of the CAT [112]. CAT is usually a carbon-loaded high-conductive tape that is wound over the main insulation in a mock stator core as well as for the core end position. SGT is a silicon carbide (SiC) powder-loaded semi-conductive tape that is partially overlapped on the end of CAT.

Partial discharges occur at the interface of CAT and SGT because the potential gradient becomes large at this area. Even if the SG coating is designed to keep the electric field below the surface discharge level, this capability will be necessarily accompanied by increased heat generation in the SGT. Degradation [113] and loss of resistivity and nonlinearity of the SG material [114] can be consequences of an increased temperature that in some cases can reach 55°C above that observed under sinusoidal 60/50 Hz frequency [115]. As a consequence, not only the electric field but also the ohmic heat generation for inverter-fed motors must be controlled [114, 115].

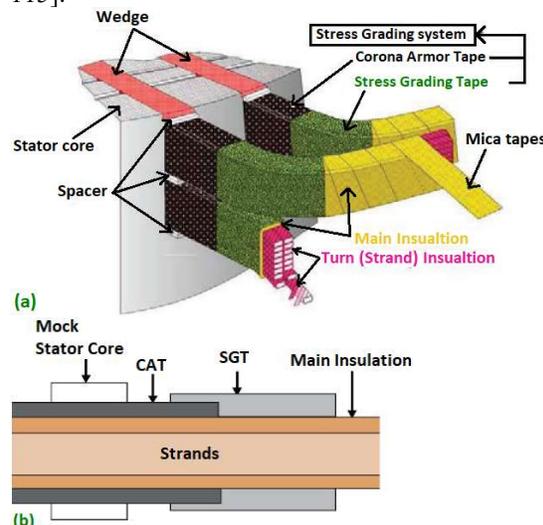

**Figure 5.** (a) Insulation system of a slot end, (b) Cross section of end-winding of stator.

However, all these studies are limited up to a few kV/μs for slew rate and a few kHz switching frequency. Studies show that both switching frequency and slew rate are critical parameters in the heat level generated on the SG system.

### 4.3 THERMAL MODELING OF CABLE TERMINATIONS

In cable terminations subjected to fast, repetitive voltage pulses, field enhancement and the flow of capACitive current cause high power loss in both the semiconducting and conductive stress grading layers, resulting in the generation of hot spots and thermal breakdown [94, 120-123]. Thermal and electro-thermal models were developed to predict the temperature distribution inside the cable termination [120, 121, 123-125], particularly under square wave pulses [121, 123] with a good agreement with test results [124], reporting increase temperatures of up to 9-10°C as shown in Figure 6 for slip-over cable terminations. Here the technical gaps in the research carried out so far are that both experiments and models have been done for up to a few kHz. In Figure 6, it can be seen that with increasing frequency, from 2 to 4 kHz, temperature rise does not linearly increase and depends on voltage amplitude. If we assume the trend shown in Figure 6 holds for up to 500 kHz, for a voltage amplitude of 12 kV, an increased temperature up to 1200°C results, which undoubtedly leads to thermal breakdown.

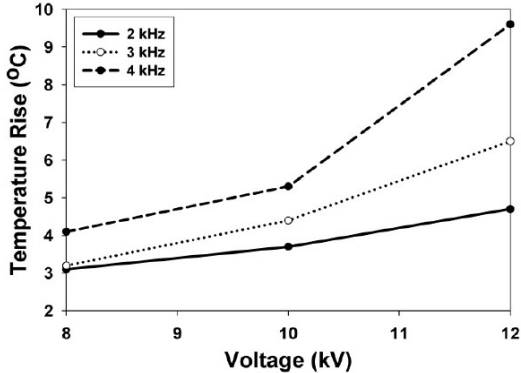

**Figure 6.** The maximum measured surface temperature rises for cable termination at different voltages and switching frequencies under square pulse [122].

### 4.4 INSULATION MODELING OF SOLID-STATE TRANSFORMERS

The basic idea of a solid-state transformer configuration is that the 50/60 Hz ac voltage is transformed into a medium frequency (MF) one in a few kHz to a few hundred kHz. Then this MF voltage is stepped up/down by an MF transformer with significantly decreased volume and weight. Finally, it is shaped back into the desired 50/60 Hz voltage to feed the load. MF voltage pulses with high $dv/dt$ originated by SiC-based converters of the SST provide a new type of electrical stress on insulation systems of the SST (both on its converters and especially on its transformer providing galvanic isolation within the SST), that is not experienced in a conventional transformer.

Through electric field calculation in a multi-cell MV (designed for a connection to the 10-kV grid) MF (the considered switching frequencies were 3 kHz for the AC-DC converter and 8 kHz for the DC-DC stages) SST, it was shown that the transformer insulation is the most stressed part [122]. The influence of the SST topology on the insulation stress was shown where a two-level legs full bridge can reduce the maximal MF RMS electric field by 35% than a three-level Neutral-Point-Clamped (NPC) bridge legs (3LL) half-bridge [126].

In [127] it was shown that the insulation losses might be used as a diagnostic tool for MV power electronic converter insulation. In this regard, it was shown that the duty cycle and the modulation index of PWM waveforms have a minor impact on the insulation losses while the voltage amplitude, the switching frequency, and the material parameters play a crucial role in the losses in the insulation.

Figure 7 shows the normalized dielectric losses, $P'$, for different rise times and switching frequencies. In this regard, the frequency plays a very powerful role while the losses increase only logarithmically for the reciprocal of the rise time. $P'$ was obtained through a closed-form analytical expression for a PWM voltage with a constant duty cycle given by $P = (\varepsilon'' C_0 V_{DC}^2)P'$, $P' \approx (2f_s/\pi) \ln[(2e^\gamma/f_s)\sin(\pi D_c)]$ where $P$ is the dielectric loss, $\varepsilon''$ is the imaginary part of the complex relative permittivity of insulating material, $\varepsilon(f)$, given by $\varepsilon(f) = \varepsilon_0(\varepsilon'(f) - j\varepsilon''(f))$, $C_0$ is the vacuum capacitance computed with $\varepsilon_0$, $D_c$ is the duty cycle, $V_{DC}$ and $f_s$ are the amplitude and switching frequency of the PWM voltage, respectively [127].

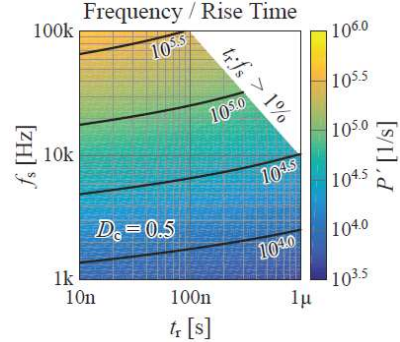

**Figure 7.** Impact of the switching frequency and the rise time on the dielectric losses [127].

Closed-form equations for dielectric loss are also developed for PWM with sinusoidal modulation and PWM with frequency-dependent materials [128]. Some approximations have been made in obtaining dielectric loss equations, thus allowing for further research to develop more precise calculations. Findings in [128] show that insulation materials working at 50/60 Hz frequency, may lead to a thermal runaway resulted from dielectric loss for working at nominal power or permissible overload under MF voltages. Further research is needed to discover novel insulating materials capable of operating under higher temperatures and higher frequencies.

High surface electric fields of 21.2 kV/cm peak were calculated for an MV, MF transformer used in an SST demonstrator [128]. As a solution, a resistive shielding preventing capacitively coupled disturbances was proposed for electric field reduction, however, it cannot suppress conducted EMI. Further research is needed to address this issue and also finding optimal shielding conductivity and type for different types of transformers and under DC/MF/HF cases.

Generally, the trends toward higher switching frequency and higher slew rates in MV voltage level, as well as compact and oil-free designs cause accelerated aging of insulation materials used in SSTs. New dielectric materials and mitigation methods, as well as accurate electric field and dielectric loss calculation techniques, should be investigated to address this issue.

## 5 CONCLUSIONS

Accelerated aging of insulation systems under fast (with a $dv/dt$ up to 100 kV/$\mu$s), repetitive (with a frequency up to 500 kHz) voltage pulses generated by next-generation WBG-based conversion systems is the most significant barrier to benefit from these conversion systems in power systems for different applications. Frequency and slew rate are two of the most important factors of a voltage pulse, influencing the level of degradation of the insulation systems. However experimental investigations carried out so far and reviewed in this paper to study their influences are limited to a few kHz and a few ten kV/$\mu$s. Moreover, tests have been done for simple electrode geometries that their results cannot be used for complex, realistic situations. Thus, Multiphysics modeling is proposed to understand the mechanisms behind pre-breakdown phenomena and as a powerful tool for electrical insulation design. Furthermore, the need to develop very high-frequency EMT models for different components under fast, repetitive voltage pulses generated by WBG-conversion systems to obtain accurate distributions of voltage and electric field on different sections of a power system was discussed. The paper provides a framework for future research in dielectrics and electrical insulation design for systems under fast, repetitive voltage pluses originated by WBG-based conversion systems.

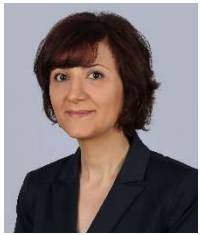

**Mona Ghassemi** (S'07-M'13-SM'16) received her M.S. and Ph.D. degrees both with the first honor in electrical engineering (power system and high voltage engineering) from the University of Tehran, Iran in 2007 and 2012, respectively. She spent two years researching as Postdoctoral Fellow at high voltage laboratory of NSERC/Hydro-Quebec/UQAC Industrial Chair on Atmospheric Icing of Power Network Equipment (CIGELE) and Canada Research Chair on Power Network Atmospheric Icing Engineering (INGIVRE), University of Quebec at Chicoutimi (UQAC), QC, Canada from 2013 to 2015. She also was Postdoctoral Fellow at the Electrical Insulation Research Center (EIRC) of Institute of Materials Science (IMS) at the University of Connecticut from 2015-2017. In 2017, Dr. Ghassemi joined the Bradley Department of Electrical and Computer Engineering at Virginia Tech (VT) as an assistant professor and a member of VT's Power and Energy Center (PEC) and Center for Power Electronics Systems (CPES). Dr. Ghassemi is a registered Professional Engineer in the Province of Ontario, Canada since 2015, Associate Editor of IET High Voltage Journal, and Associate Editor of International Journal of Electrical Engineering Education. She published 51 refereed journal and conference papers and one book chapter. Her research interests include electrical insulation materials and systems, high field technology, transmission line design, multiphysics modeling, electromagnetic transients in power systems and power system analysis and modeling.